\documentclass{emulateapj}

\usepackage[export]{adjustbox}

\usepackage{color}

\usepackage{soul}   

\begin{document}
\title{KIM\,3: AN ULTRA-FAINT STAR CLUSTER IN THE CONSTELLATION OF CENTAURUS}
\author{Dongwon Kim} 
\author{Helmut Jerjen}
\author{Dougal Mackey}  
\author{Gary S. Da Costa}
\author{Antonino P. Milone}
\affil{Research School of Astronomy and Astrophysics, Australian National University, Canberra, ACT 2611, Australia}

\email{dongwon.kim@anu.edu.au}

\begin{abstract}
We report the discovery of an ultra-faint star cluster in the constellation of Centaurus. This new stellar system, Kim 3,
features a half light radius of $r_{h}=2.29^{+1.28}_{-0.52}$\,pc and a total luminosity of $M_{V}=+0.7\pm0.3$. Approximately
26 stars are identified as candidate member stars down to four magnitudes below the main-sequence turn-off, 
which makes Kim\,3 the least luminous star cluster known to date.
The compact physical size and extreme low luminosity place it close to faint star clusters in the size-luminosity plane. The stellar 
population of Kim 3 appears to be relatively young ($9.5^{+3.0}_{-1.7}$\,Gyr) and metal-poor ([Fe/H]$=-1.6^{+0.45}_{-0.30}$) at a 
heliocentric distance of $15.14^{+1.00}_{-0.28}$\,kpc. The cluster lacks a well-defined center and a small but prominent group of stars consistent with the Kim 3 isochrone is present approximately 9.7 pc in projection south of the cluster center. Both are signs of the cluster being in the final stage of tidal disruption.
\end{abstract}

\section{Introduction}

The Sloan Digital Sky Survey~\citep[SDSS;][]{York2000} has unveiled a significant number of 
ultra-faint dwarf galaxies in the Milky Way halo~\citep[e.g.][]{Willman2005,Zucker2006,Belokurov2006,Irwin2007,Walsh2007,Kim2015b}. However, only a small number of new star clusters were 
found~\citep{Koposov2007,Belokurov2010,Balbinot2013,KimJerjen2015a}, with heliocentric distances 17--50\,kpc. These star clusters share both small physical sizes and low luminosities, properties
considered to be the consequences of stellar mass loss owing to internal (e.g.~dynamic relaxation) and/or 
external~(e.g. tidal stripping, tidal shocking) dynamical evolution processes \citep{Gnedin1997,Rosenberg1998}. 
This picture of the low-luminosity star clusters being strongly dynamically evolved is supported by growing observational evidence such as the presence of extra-tidal stars, flat luminosity functions, and substantial mass segregation~\citep{Carraro2007,Carraro2009,Niederste2010,Fadely2011,KimJerjen2015a,Kim2015a}.

Since the success of SDSS, other blind imaging surveys have continued searching for new stellar systems in the Milky Way halo; 
the Dark Energy Survey~\cite[DES;][]{DES}, Pan-STARRS 3$\pi$ survey (K. Chambers et al., in preparation), 
VST ATLAS survey~\citep{Shanks2015}, the Stromlo Milky Way Satellite (SMS) survey~\citep{SMS}, 
and the Survey of the Magellanic Stellar History (SMASH; D. Nidever et al., in preparation). These 
efforts have uncovered more than 20 new satellite candidates up to the present time~\citep{ Bechtol2015, Belokurov2014, 
DWagner2015, Kim2015a,KimJerjen2015b,Koposov2015a,Laevens2014,Laevens2015a,Laevens2015b,Luque2015,
Martin2015a,Torrealba2016}. Spectroscopic follow-up has revealed the kinematic and chemical characteristics of some of these systems, clarifying their nature~\citep[e.g.][]{Simon2015,Walker2015a,Walker2015b,Koposov2015b,Kirby2015b,Martin2015a,Martin2015b}. 
These ultra-faint stellar systems are rapidly filling the gap between star clusters and dwarf galaxies in the size-luminosity plane, 
rendering this diagnostic tool less effective~\citep[e.g. see discussions in ][]{Laevens2014,Belokurov2014}. 
Hence, deeper imaging and spectroscopic follow-up are becoming imperative 
to determine the true nature of the systems and possibly identify star clusters among the 
new candidates \citep[e.g.][]{Kirby2015a,Weisz2015}

In this paper we announce the discovery of a new ultra-faint star cluster, which we designate as Kim\,3, 
found in the constellation of Centaurus. This concentration of stars was detected as part of 
our ongoing imaging survey with the Dark Energy Camera (DECam) on the 4m Blanco telescope 
at Cerro Tololo in Chile. Section\,2 describes the observations and data reduction process, 
including photometry and star/galaxy separation that led to the discovery of Kim\,3. 
We also discuss the photometric calibration and completeness tests.
Section\,3 contains our analysis of the CMD and describes how we derived the properties of the 
new star cluster such as age, metallicity, distance, luminosity and structure. 
We discuss the results and draw our conclusions in section\,4.

\section{Observation, Data Reduction, and Discovery}

\begin{figure}
\begin{center}
\includegraphics[scale=0.3]{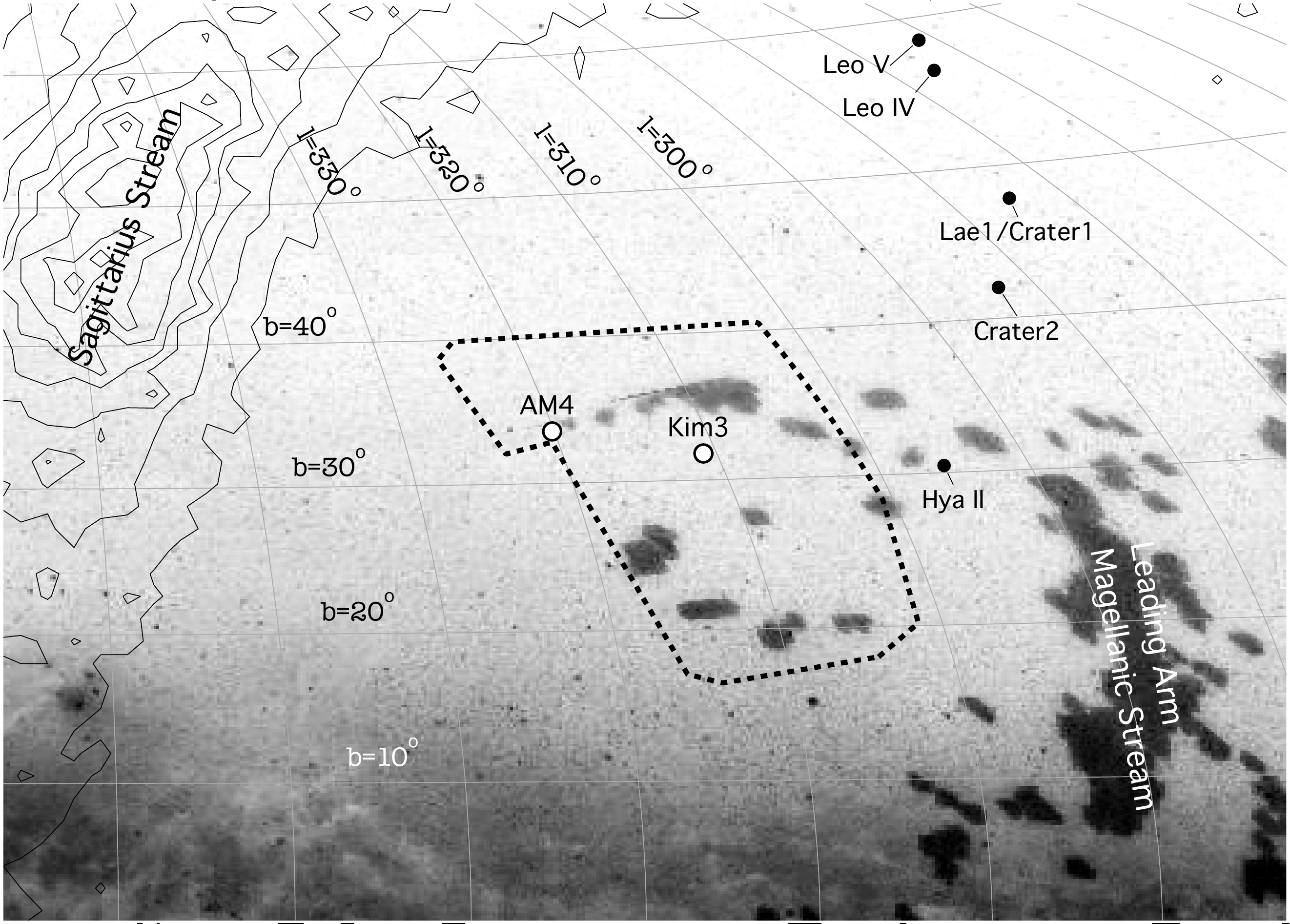}
\end{center}
\caption{Section of the sky (dashed line), showing the footprint of the $\sim 500$ square 
degree area searched for new ultra-faint Milky Way stellar systems as part of the Stromlo 
Milky Way Satellite Survey. The Galactic longitude and latitude lines are spaced 
by $10^\circ$. Known satellite galaxies and star clusters in the area are labelled. 
Black contour lines indicate particle densities for the simulated Sagittarius 
stream from~\cite{LawMajewski2010}. The background image, showing the 
end of a leading arm of the Magellanic Stream, is by~\cite{Nidever2010}, NRAO/AUI/NSF 
and Meilinger, Leiden-Argentine-Bonn Survey, Parkes Observatory, Westerbork Observatory, Arecibo 
Observatory (see http://www.nrao.edu/pr/2010/magstream/).}
\label{fig:SurveyFootprints}
\end{figure}

\begin{figure}
\begin{center}
\includegraphics[scale=0.38]{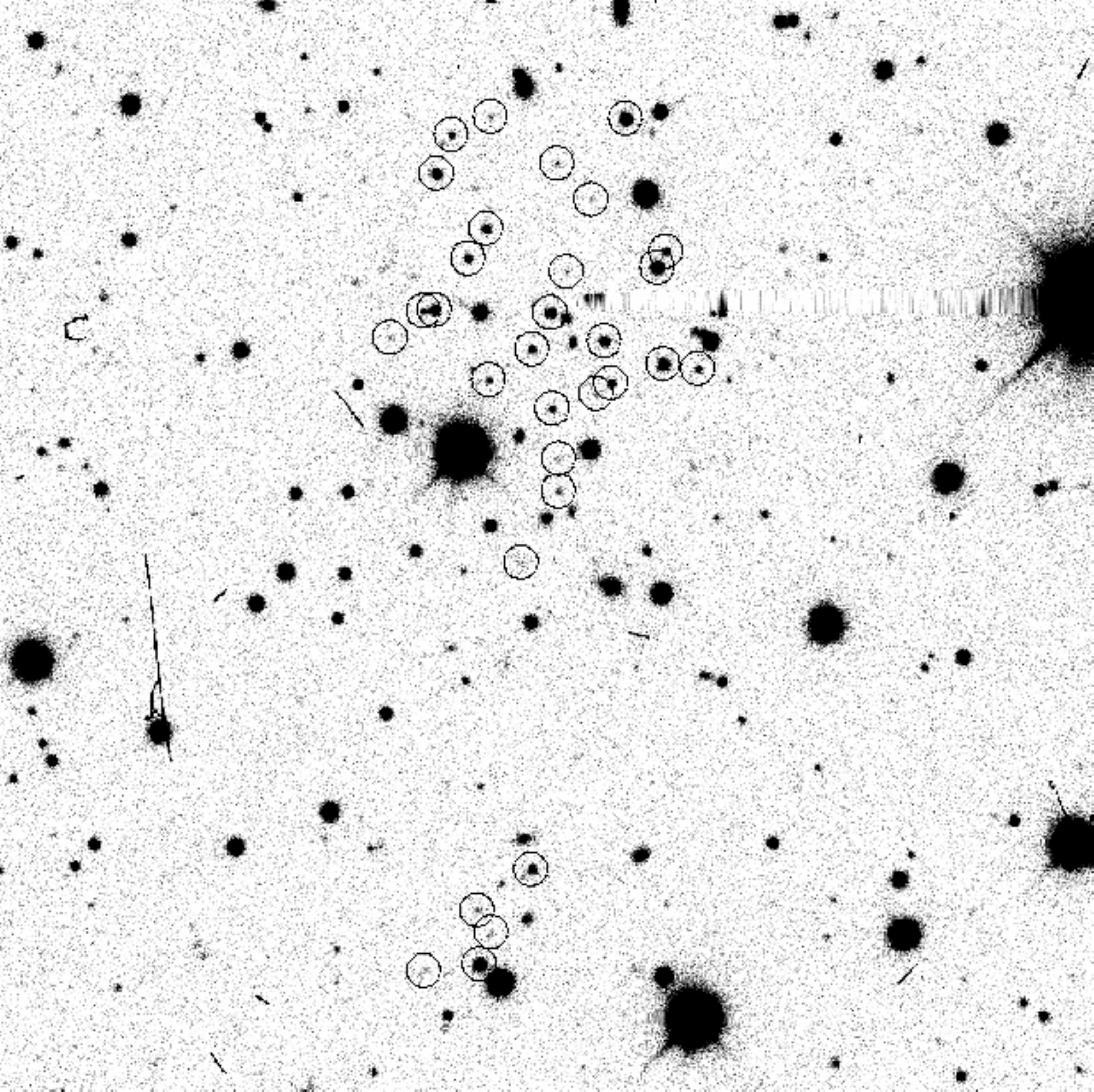}
\end{center}
\caption{$4\times 4$\,arcmin$^2$ DECam cutout $r$-band image of Kim\,3 with 210 seconds exposure time. North is up, east is to the left. Circled are all stars fainter than $r_{0}=18.5$ that are within two half-light radii from the adopted center of Kim\,3 or within a radius of $0\farcm15$ from the adopted center of the small group of stars to the south of Kim\,3 (see Figure~\ref{fig:Contour}), and consistent with the best-fitting isochrone (9.5 Gyr, [Fe/H]$=-1.6$) at a distance of 15.14\,kpc.}
\label{fig:Fits-r}
\end{figure}

\begin{figure*}
\begin{centering}
\includegraphics[scale=0.6]{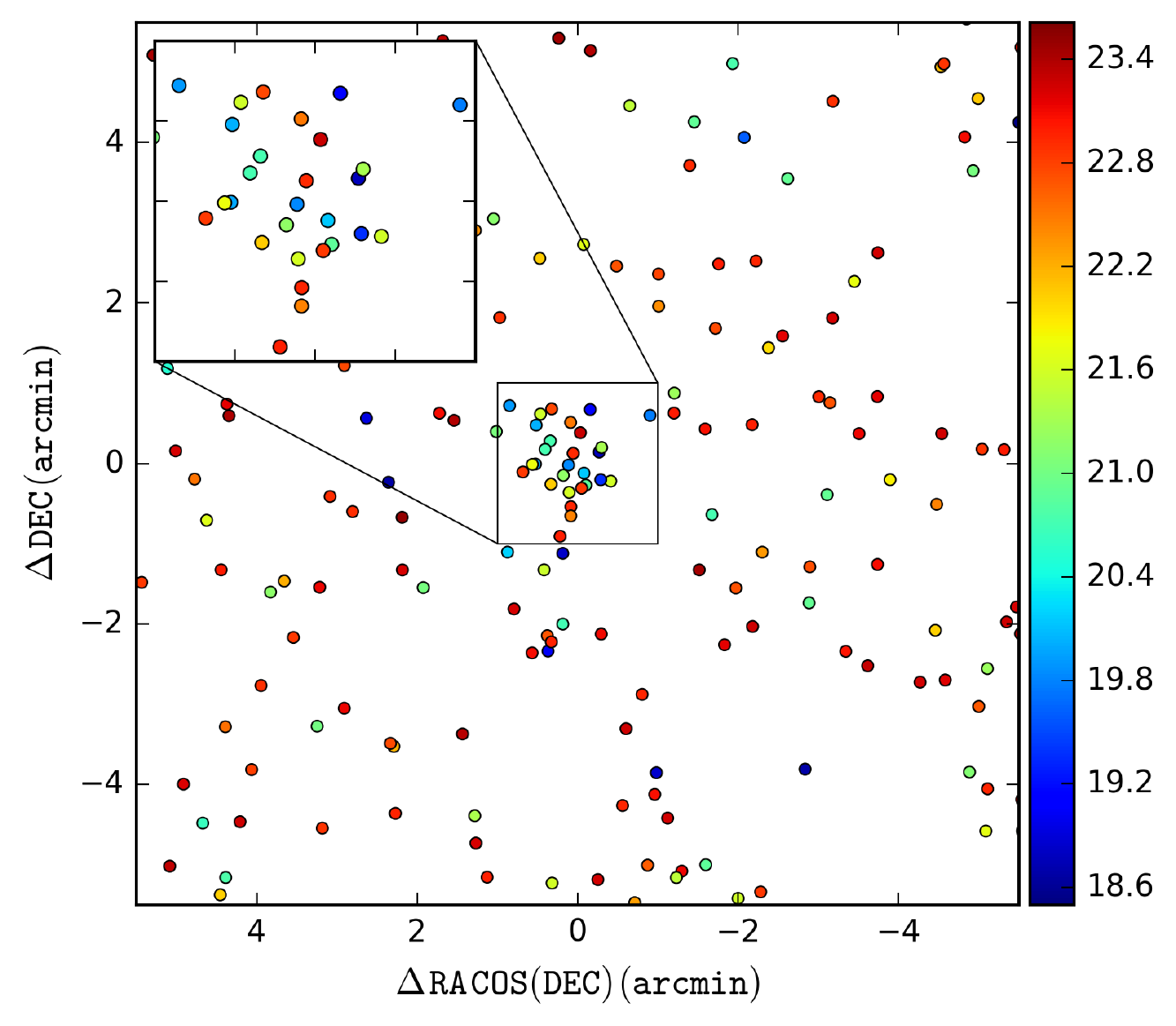}\includegraphics[scale=0.6]{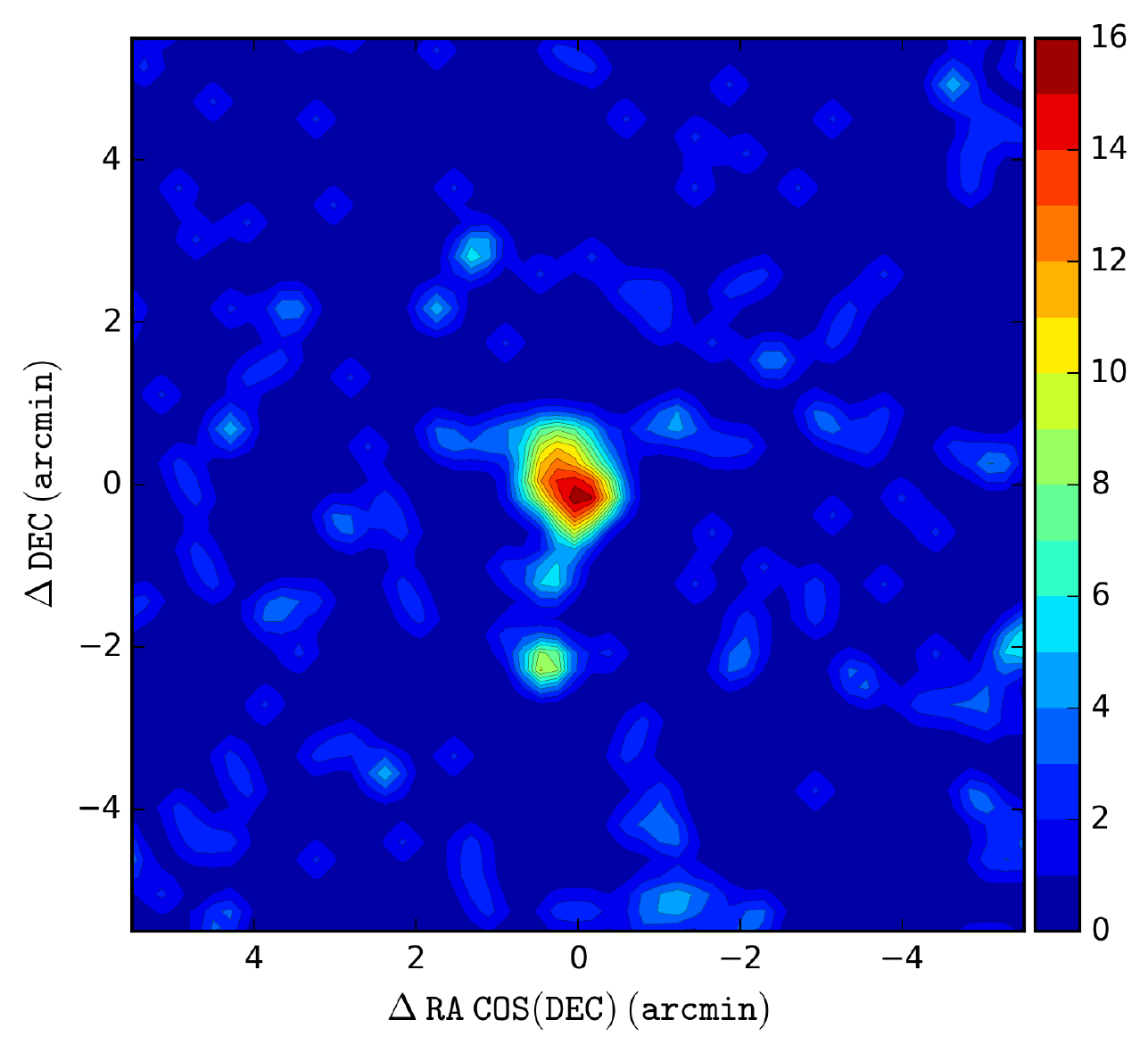}
\par\end{centering}
\caption{Left panel: distribution of candidate stars that passed the photometric filtering process centred on Kim\,3 in the $11\times11$ square arcmin window. Right panel: smoothed density contour map corresponding to the left panel. The contour levels mark the stellar density in units of the standard deviation above the background. }
\label{fig:Contour}
\end{figure*}

As part of the Stromlo Milky Way Satellite (SMS) Survey project we have observed in non-targeted mode $\sim500$ square degrees of sky in the Centaurus region (see Figure~\ref{fig:SurveyFootprints}) using the Dark Energy Camera~\citep[DECam; ][]{DECam} of the 4-m Blanco Telescope located at Cerro Tololo Inter-American Observatory (CTIO). The imager consists of 62 2k $\times$ 4k CCD chips with a pixel scale of $0\farcs27$, which delivers a $\sim3$ square degree field of view. We obtained images in the $g$ and $r$ bands over two observing runs in July 2014 and June 2015 as part of observing proposals 2014A-0624 and 2015A-0616 (both PI: H.~Jerjen). More details on the former observing run can be found in our previous work \citep{Kim2015a}. In the case of the latter session, we set the exposure times to between 100 and 210\,s depending on the fraction of moon illumination and the angular distance of the target field from the moon. To fill the inter-chip gaps, we dithered in a diagonal direction by half of a single chip in both $x$ and $y$ for each field, providing two exposures per field per filter. The images were reduced using the DECam community pipeline~\citep{DECamCP2014}. This process includes bias subtraction, dark and flat-field corrections, and the application of a WCS solution to each image.

We carried out PSF photometry over the pre-processed single exposure images to produce photometric catalogs using SExtractor/PSFEx~\citep{SExtractor,PSFEx} on a local 16 node/128 core computer cluster. For the star/galaxy separation we made use of the
$\mathtt{SPREAD\_MODEL}$ parameter provided by SExtractor, where the threshold was set $\mathtt{\left| SPREAD\_MODEL\right|<0.003+SPREADERR\_MODEL}$ as described in \cite{Koposov2015a}. This process was applied to the photometric band that exhibited the better defined PSF over the entire field. The $g$ and $r$ band catalogs were crossmatched using STILTS~\citep{STILTS} with a $1\arcsec$ tolerance. The instrumental magnitudes were then calibrated with respect to the APASS DR\,8\footnote{https://www.aavso.org/apass} star catalog via bootstrap sampling with 500 iterations and 3-sigma clipping. The number of matched stars in a field ranged between 100--1600. Finally, each calibrated object was corrected for Galactic extinction based on the reddening map by~\cite{Schlegel1998} and the correction coefficients from~\cite{Schlafly2011}.

We ran our overdensity detection algorithm, which is based on the method of \citet{Invisibles}, over the final point source catalogue that was produced for each field by our photometry pipeline. For more details about the algorithm see \cite{KimJerjen2015a}. Briefly, the algorithm enhances the contrast between satellite population and the Milky Way foreground stars by using photometric filters in the color-magnitude space and comparing the integrated signal-to-noise ratios (SNRs) of point-source clusters on a convolved stellar density map in the field of view of DECam. In this search, we recovered the known globular cluster AM4 and detected the new object Kim\,3, the SNR of which reached the $10\sigma$ level over the Poisson noise measured in the surrounding point-source distribution.

\begin{figure*}[t!]
\begin{centering}
\includegraphics[scale=0.85]{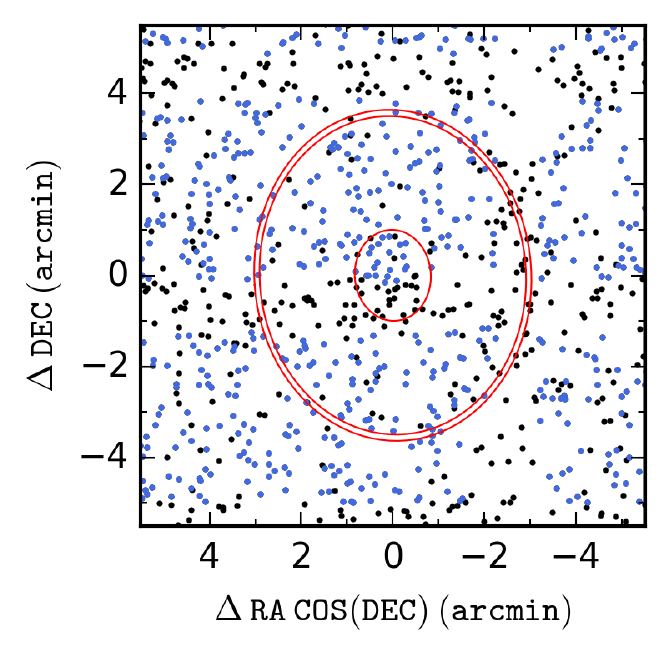}\includegraphics[scale=0.85]{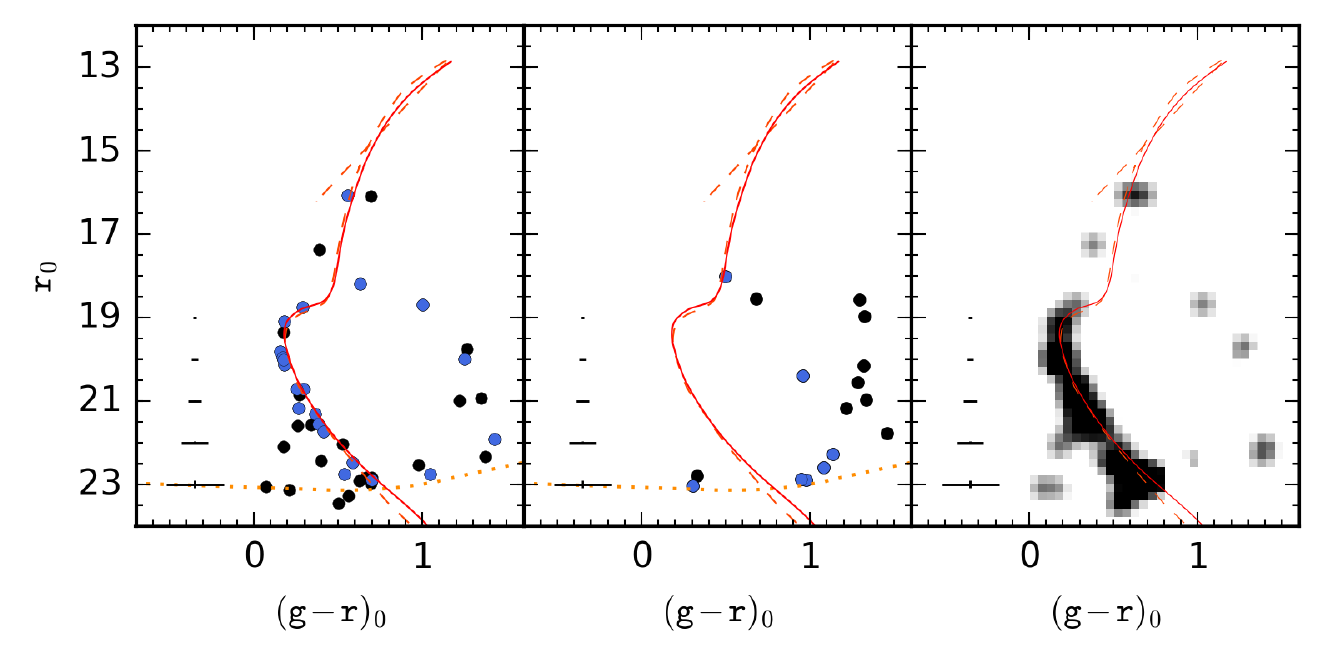}
\includegraphics[scale=0.85]{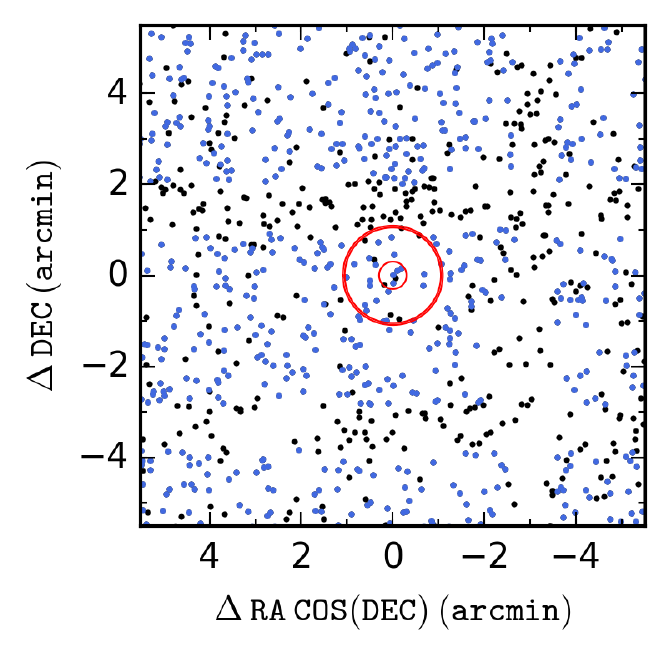}\includegraphics[scale=0.85]{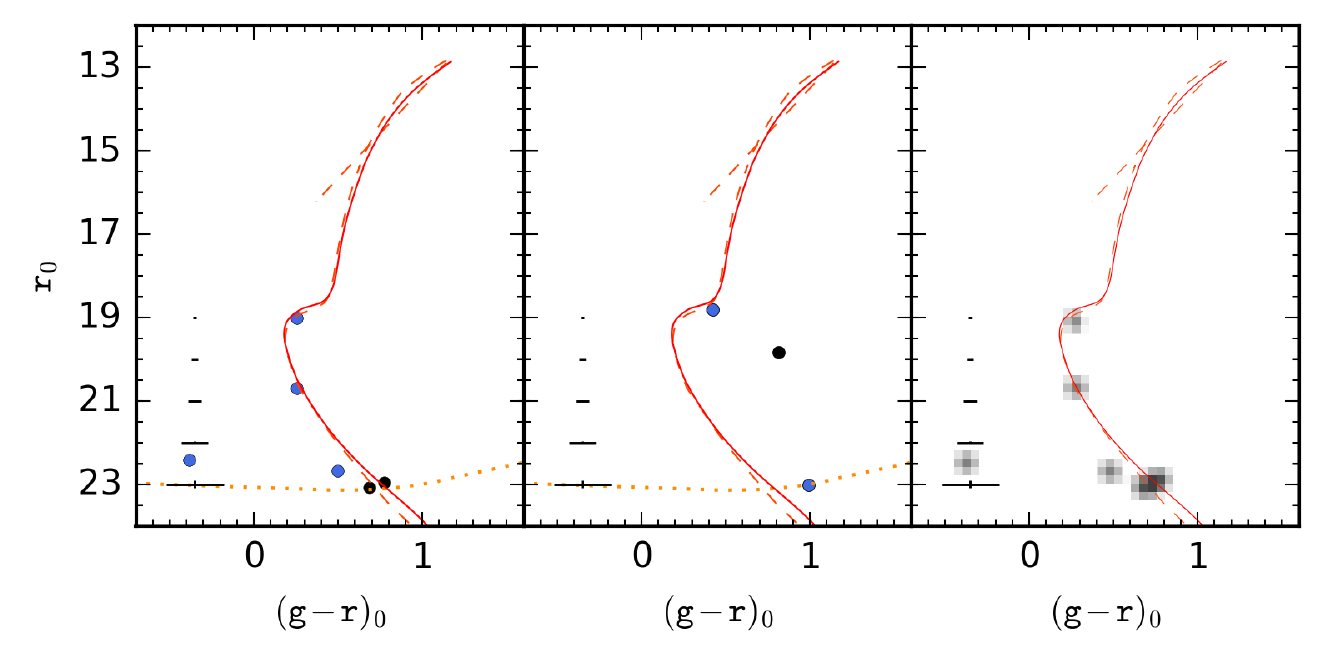}
\par\end{centering}
\caption{Upper left panel: distribution of all stars in the same window as Figure~\ref{fig:Contour}. The blue markers represent stars detected in all four exposures taken for the field and thus have two independent photometric measurements per filter. Black markers are stars with one measurement per filter. The three ellipses mark elliptical radii of $1\farcm00$ ($\sim 2r_{h}$), $3\farcm50$($\sim 7r_{h}$) and $3\farcm54$, respectively, with a position angle of $4$ degrees and an ellipticity of 0.17. Upper middle left panel: CMD of stars lying in the inner ellipse of the left panel. Overplotted are the best-fitting Dartmouth (red solid line) isochrone of age 9.5 Gyr and [Fe/H]$=-1.6$ and the PARSEC (red dashed line) isochrone of the same age and metallicity for comparison, shifted to the distance modulus of $(m-M)=15.90$. The magnitudes of the stars with two measurements (blue markers) were uncertainty-weighted averaged so that their final uncertainties are $\sim30$\% smaller than the stars with one measurement. The dotted line indicates the 50\% completeness level of our photometry. Upper middle right panel: control CMD of stars that reside between the two outer ellipses. Upper right panel: the differential Hess diagram, the inner CMD minus the control CMD, for which each CMD was binned onto a grid with intervals of 0.05 mag for $(g-r)_{0}$ and 0.2 mag for $r_{0}$. Lower panels: the same as the upper panels but for the small overdensity $\sim2\farcm2$ south of Kim\,3 shown in Figure~\ref{fig:Contour}. Its central coordinates (J2000) were visually determined as (RA, DEC)=($200.696^{\circ}$, $-30.637^{\circ}$).  Its radius was measured $\sim0\farcm15$. The three circles mark radii of $0\farcm30$, $1\farcm05$ and $1\farcm09$. \label{fig:CMD}}
\end{figure*}

We performed completeness tests for the photometry as follows. We first created an accurate PSF model image using the PSF task of DAOPHOT in the IRAF environment and then added 100 artificial stars per chip at random pixel coordinates using the ADDSTAR task in IRAF. A series of images were produced for different input magnitudes at 0.5\,mag intervals. We then ran our photometry routine and measured the recovery rate, for which we also applied the same star/galaxy separation criteria for more 
realistic measurements. This procedure was repeated 20 times to obtain reliable statistics. The completeness function for our CMD was then finalised by multiplying the recovery rates in the $g$ and $r$ bands as the two catalogues were cross-matched to generate the CMD. The  90\%  and 50\% levels of our photometry at the color $(g-r)=0.5$ correspond to $r=20.74$ and $r=23.21$ respectively. The 50\% completeness level as a function of color and $r$ magnitude is indicated by the dotted lines in Figure~\ref{fig:CMD}.

Figure~\ref{fig:Fits-r} shows a $r$-band cutout image centred on Kim\,3, where the cluster is completely resolved into individual stars. We note that a very bright star to the west of the cluster caused a ``blooming'' effect across the image, which was automatically corrected via linear interpolation by the NOAO community pipeline. It is possible that some Kim\,3 member stars are hidden behind the interpolated region. Another bright star to the south-east of the nominal cluster center could also be hiding stars associated with the cluster.

The left panel of Figure~\ref{fig:Contour} shows the R.A.-DEC distribution of all stars that passed the photometric filter based on the main-sequence and its turn off of the best-fitting isochrone. The color of the marker represents the $r_{0}$ magnitude. The right panel 
shows the corresponding convolved density map where the contours are in units of the measured standard deviation. 
The contours reveal an asymmetric feature within high density level isophotes ($>6\sigma$) and a tail-like structure to the south, 
which could be evidence of tidal disruption. We note that a similar feature has been found in the case of the dissolving star cluster 
Kim\,1 \citep[see Figure 2 in][]{KimJerjen2015a}. 
At the end of that structure a small, but prominent overdensity clearly stands out $\sim2\farcm2$ below 
the center of Kim\,3. This feature further strengthens the impression that the tail is actually the product of the 
disruption process in the cluster.

\begin{deluxetable}{lrl}
\tablewidth{0pt}
\tablecaption{Properties of Kim\,3}
\tablehead{
\colhead{Parameter} & 
\colhead{Value} &
\colhead{Unit}}
\startdata
$\alpha_{J2000}$ & 13 22 45.2$\pm$2.0 & h m s \\
$\delta_{J2000}$ & -30 36 03.6$\pm$2.0 & $^\circ$ $\arcmin$ $\arcsec$ \\
$l$ & 310.860 & deg\\
$b$ & 31.788 & deg\\
$(m-M)$ & $15.90^{+0.11}_{-0.04}$ & mag \\
$ $[Fe/H] & $-1.6^{+0.45}_{-0.30}$ & dex\\
$d_\odot$ &  $15.14^{+1.00}_{-0.28}$ & kpc \\
$d_{gal}$ & $12.58^{+0.85}_{-0.23}$ & kpc \\
$r_{h}$ & $0.52^{+0.24}_{-0.11}$ & \arcmin \\
$r_{h}$ & $2.29^{+1.28}_{-0.52}$ & pc \\
$\epsilon$ & $0.17^{+0.26}_{-0.17}$ & \\
$\theta$ & $4\pm24$ & deg \\
$M_{V}$& $+0.7\pm0.3$ & mag \\
$E(B-V)$\tablenotemark{a}& 0.061 & mag 
\enddata
\tablenotetext{a}{From~\cite{Schlafly2011}.}
\label{tab:Parameters}
\end{deluxetable}

\section{Kim\,3 Properties}

\subsection{Color-Magnitude Diagram}
The left panel of Figure~\ref{fig:CMD} shows all star-like objects from our analysis found in the vicinity of Kim\,3 where blue markers represent objects in common in all four exposures, $2\times100$\,s in $g$ band and $2\times210$\,s in $r$ band. The next three panels on the right correspond to the CMD of stars within two half-light radii of Kim\,3, the control CMD and the differential Hess diagram. We calculated the uncertainty weighted average magnitudes for the overlapping objects so that their photometric uncertainties are $\sim30$\% smaller than those with the single measurements. The CMD of Kim\,3 possesses stars over $\sim4$ magnitudes, from the MSTO down to the 50\% completeness level, that are consistent with an old (9.5 Gyr) and metal-poor ([Fe/H] $=-1.6$) population at a distance modulus of $(m-M)=15.90$. Such a tight main-sequence fit has also been noticed in the CMD of Kim\,1. The CMD of the smaller overdensity nearby Kim\,3 also shows a fairly consistent fit to the same isochrone within photometric uncertainties. Its true affilliation to Kim\,3 can be determined once spectroscopic data become available.

\subsection{Age, Metallicity and Distance Modulus}
We estimate the age, metallicity, and distance of Kim\,3 using the maximum likelihood method described 
in~\cite{Frayn2002}, \cite{Fadely2011}, and \cite{KimJerjen2015a}. For the analysis we use all stars within an elliptical radius 
of $1\farcm0$ from Kim\,3, the inner ellipse in the upper left panel of Figure~\ref{fig:CMD}.  We calculate the 
maximum likelihood values as defined by the Equations\,1 and 2 in Fadely et al. (2011), over a grid 
of Dartmouth model isochrones \citep{Dartmouth}. The grid points in the multi-dimensional parameter 
space cover the age range from 7.0--13.5\,Gyr, a metallicity range $-2.5\leq$ [Fe/H] $\leq-0.5$\,dex, and a distance range $15.7<(m-M)<16.3$.  Grid steps are 0.5\,Gyr, 0.1\,dex, and 0.05\,mag, respectively.

Due to the small sample size relative to the number of free parameters, we chose to fix [$\alpha$/Fe] in the fit to ensure adequate convergence of the ML algorithm. We tested two scenarios $-$ the first with [$\alpha$/Fe]=+0.4 to match most known Galactic globular clusters, and the second with [$\alpha$/Fe]=0.0 to match the small sample of younger globular clusters seen in the MW halo~\citep[e.g.][]{Cohen2004,Sbordone2005,Sakari2011,Villanova2013}. We found a similar age and distance modulus for each scenario, but rather different values of metallicity: age$=9.5^{+1.8}_{-1.0}$\,Gyr, $(m-M)=15.93^{+0.11}_{-0.03}$\,mag, and [Fe/H]$=-2.0^{+0.35}_{-0.40}$ for [$\alpha$/Fe]=+0.4, and $9.5^{+3.0}_{-1.7}$\,Gyr, $(m-M)=15.90^{+0.11}_{-0.04}$\,mag, and [Fe/H]$=-1.6^{+0.45}_{-0.30}$ for [$\alpha$/Fe]=0.0. With the first solution Kim\,3 would be a significant outlier in the age-metallicity relationship observed for Galactic globular clusters~\citep[see Figure 10 in][]{Dotter2011}, but with the second it would agree much more closely. Given that Kim\,3 appears to have a relatively young age, we adopt the solution for [$\alpha$/Fe]=0.0 as our final estimate for the rest of the paper.  However, we will ultimately need spectroscopic follow-up to confirm these results.

In Figure~\ref{fig:MLplot}, we present the matrix of likelihood values for the sample described above 
after interpolation and smoothing over two grid points. The best-fitting Dartmouth isochrone (red 
solid line in Figure~\ref{fig:CMD}) has an age of 9.5 Gyr, [Fe/H] $= -1.6$\,dex, $[\alpha/$Fe$]=0.0$ 
with a heliocentric distance of 15.14\,kpc ($m-M=15.90$\,mag). These estimates also yield a consistent fit for the PARSEC model isochrones (red dashed line in Figure~\ref{fig:CMD}). The 68\%, 95\%, and 99\% confidence 
contours are overplotted in Figure~\ref{fig:MLplot}.  

\subsection{Structural Parameters and Luminosity}

\begin{figure}
\includegraphics[scale=0.46]{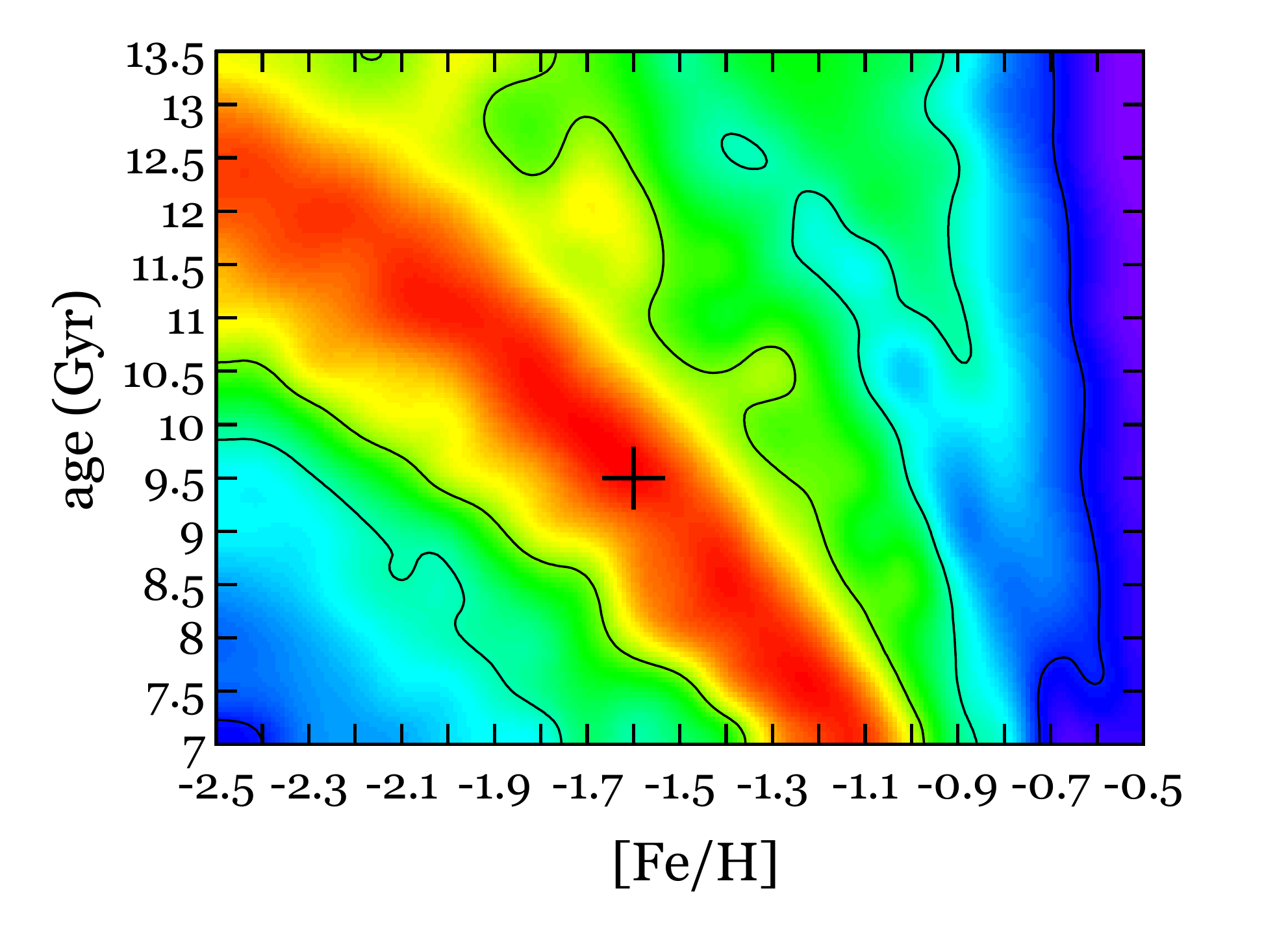}
\caption{
Smoothed maximum likelihood density map in age-metallicity space for all stars within 2$r_{h}$ around Kim\,3. 
Contour lines show the 68\%, 95\%, and 99\% confidence levels. The diagonal flow of the contour lines reflects the
age-metallicity degeneracy inherent to such an isochrone fitting procedure. The 1D marginalized parameters 
around the best fit with uncertainties are listed in Table\,1.
}
\label{fig:MLplot}
\end{figure}

\begin{figure}[t!]
\begin{centering}
\includegraphics[scale=0.83]{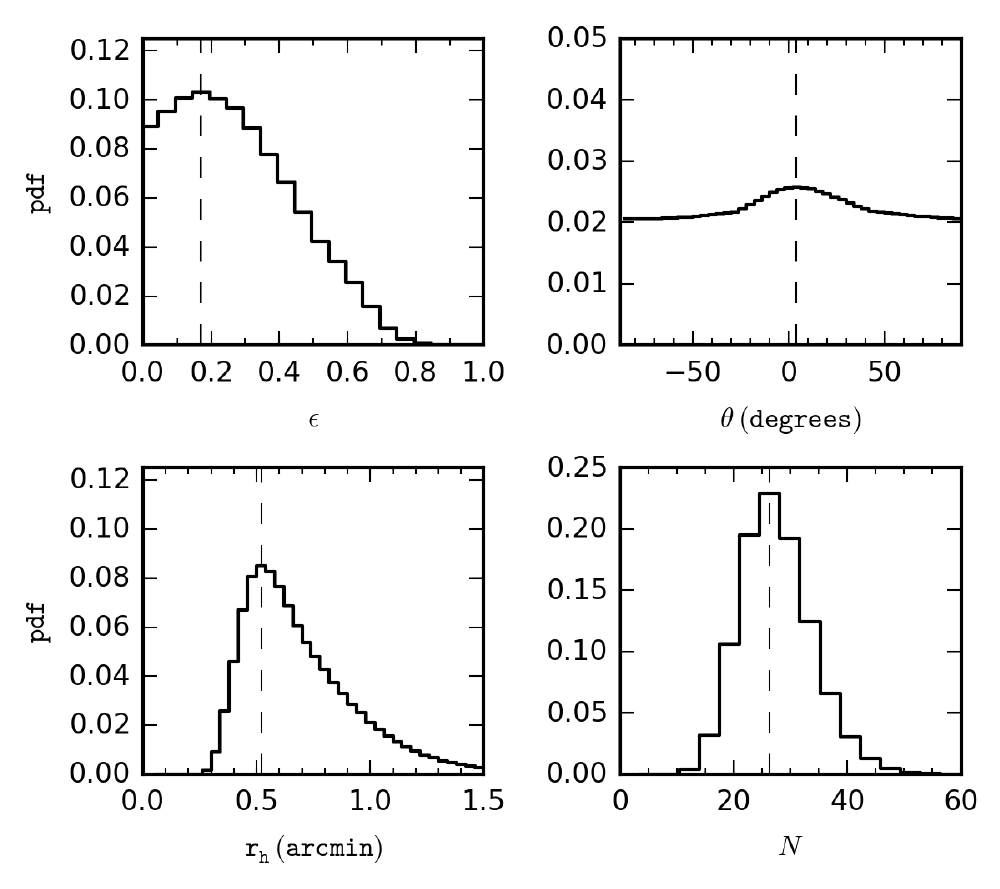}
\includegraphics[scale=0.79]{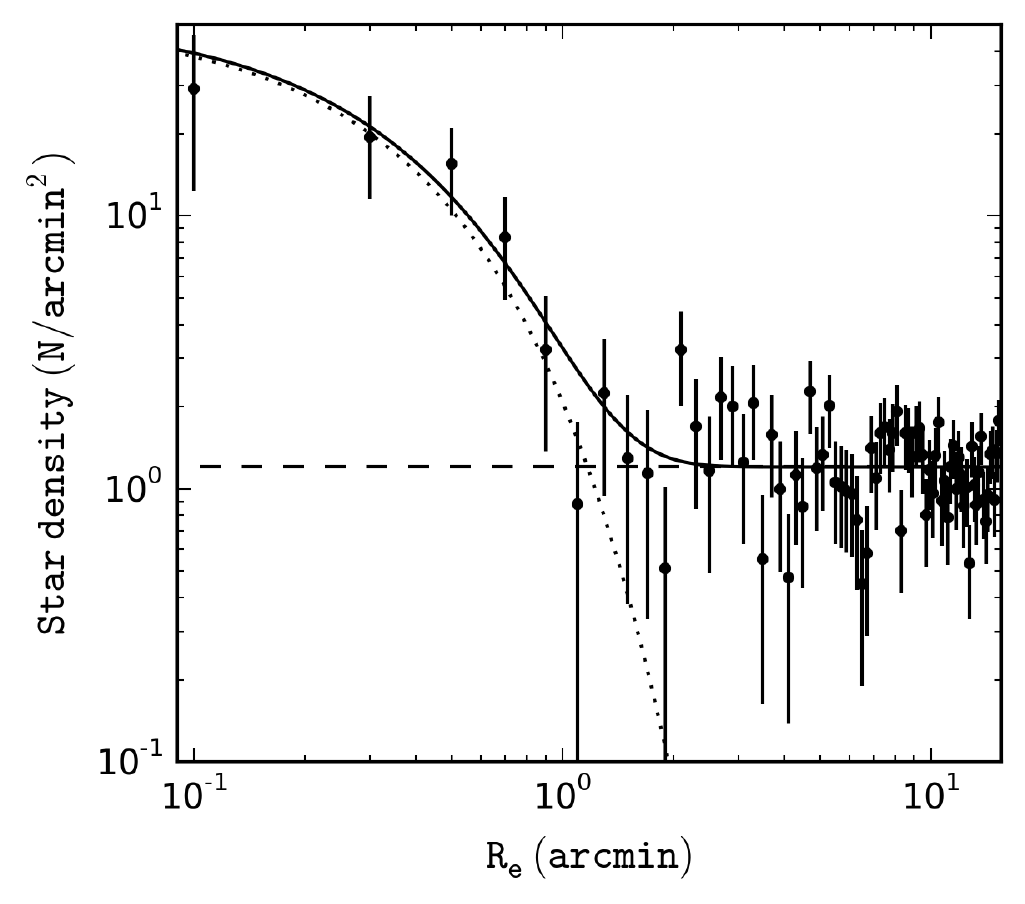}
\par\end{centering}
\protect\caption{Upper panels: marginalized probability distribution functions for the structural parameters of Kim\,3 : the ellipticity  ($\epsilon$),  the position angle ($\theta$), half-light radius ($r_{h}$) and the number of stars that belong to the cluster in our photometry ($N$). The 
mode for each parameter is marked by a vertical dashed line. Bottom panel: radial density profile of Kim\,3 based on mode values as a function of elliptical radius $R_{e}$. The dotted, dashed and solid lines correspond to the best-fit exponential model, the foreground level and the combined fit respectively. \label{fig:RadialProfile}}
\end{figure}

\begin{figure}
\begin{center}
\includegraphics[scale=0.65]{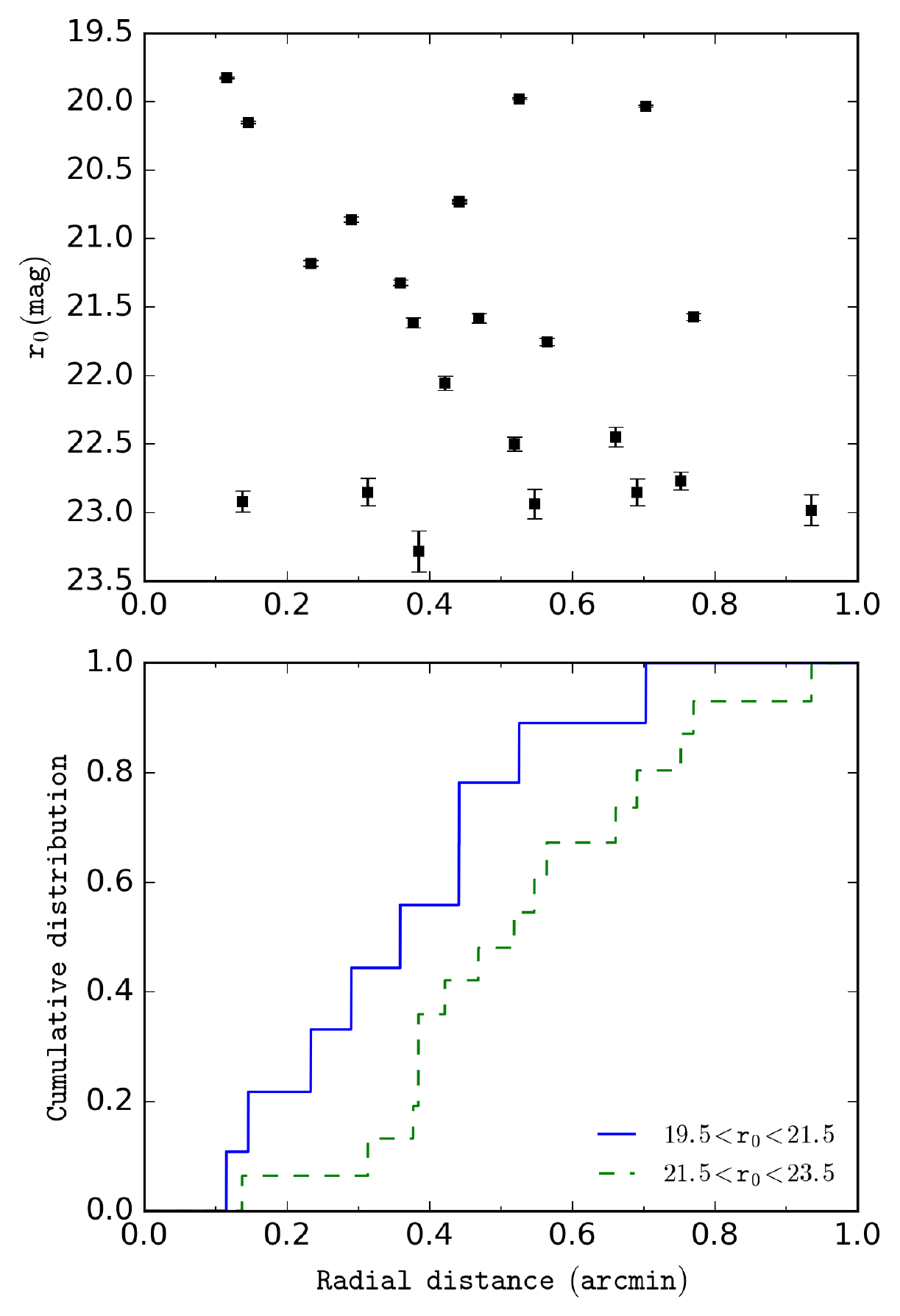}
\end{center}
\caption{ Upper panel: $r_{0}$ magnitudes of all Kim\,3 main sequence stars within $2r_{h}$($\sim1\farcm0$) as a function of radial distances from the center of the cluster. Lower panel: completeness-corrected cumulative distribution functions for two different magnitude intervals.\label{fig:MassSeg}}
\end{figure}

To determine the central coordinates and structural parameters of Kim\,3, we employed the Maximum Likelihood (ML) 
routine introduced in \cite{Martin2008} using the stars fainter than $r_{0}=18.5$ mag that passed the photometric 
filtering process. The upper panels of Figure~\ref{fig:RadialProfile} show the resulting marginal probability distribution 
functions (pdfs) for the structural parameters and the bottom panel shows the radial density profile with the 
exponential profile using the modal values from the ML analysis. Formally, Kim 3 is mildly elliptical with $\epsilon=0.17^{+0.26}_{-0.17}$ at position angle $\theta=4\pm24$ degrees; however as is evident from the pdfs in Figure~\ref{fig:MLplot}, these quantities are not well constrained by the available data. The physical half-light radius of the 
cluster is calculated as $r_{h}=2.29^{+1.28}_{-0.52}$\,pc adopting the heliocentric distance 
of $15.14^{+1.00}_{-0.28}$ kpc derived in Section 3.2. This shows that Kim\,3 is similar in size to 
Segue 3~\citep{Fadely2011}. We note that the exclusion of possible member stars obscured by the blooming 
effect or the bright star near the cluster (see Figure~\ref{fig:Fits-r}) could slightly affect the results. The number of 
stars that belong to the cluster $N$ was calculated with eq. (5) in~\cite{Martin2008}. 

We estimated the total luminosity of Kim\,3 using the star count parameter $N$ as follows. We first multiplied the
normalised theoretical luminosity function (LF) with the completeness function determined in Section 2. 
We then integrated the LF as a probability density function of magnitude. The ratio of the star count 
mode $N=26$ to the probability density gives the scale factor to transform the original LF to the observed level. 
Finally we calculated the weighted integral of flux treating the scaled LF as a weight function. We obtained $M_{r}=+0.51+^{0.27}_{-0.29}$ using the Dartmouth luminosity function of 9.5 Gyr and [Fe/H]$=-1.6$  with the mass function by \cite{Chabrier2001} and $M_{r}=+0.43+^{0.28}_{-0.30}$ using the PARSEC luminosity function of the same age and metallicity with the mass function by \cite{Kroupa2001}. The quoted errors include the uncertainties in the star count parameter $N$ and 
the distance modulus derived in Section 3.2. For a 9.5 Gyr and [Fe/H]=-1.6 stellar population, the Dartmouth and PARSEC models have a mean color $V-r=0.21$ and $V-r=0.22$ respectively, which convert both the $M_{r}$ magnitudes into $M_{V}=+0.7\pm0.3$. We adopted $M_{V}=+0.7\pm0.3$ as our final estimate of the total luminosity of Kim\,3. All derived parameters presented in this section are 
summarized in Table~\ref{tab:Parameters}.

\section{Discussion and Conclusion}

We report the discovery of the ultra-faint star cluster Kim\,3 in the constellation of Centaurus. 
It is a compact ($r_{h}=2.29^{+1.28}_{-0.52}$) and extremely faint ($M_{V}=+0.7\pm0.3$) star cluster. Although its physical size and ellipticity are comparable to Segue\,3, the new 
cluster exhibits even more asymmetry on the contour map than Segue\,3 \citep[see Figure 5 
in][]{Fadely2011}. Kim\,3 is even slightly fainter than Kim\,1~\citep[$M_{V}=+0.3\pm0.5$;][]{KimJerjen2015a}, 
Segue\,3~\citep[$M_{V}=+0.0\pm0.8$;][]{Fadely2011} and Mu{\~n}oz\,1~\citep[$M_{V}=-0.4\pm0.9$][]{Munoz2012} 
and thus sets a new record in the size luminosity plane. The best-fitting model isochrone in the CMD indicates 
that the stars of Kim\,3 are located at a heliocentric distance of $15.14^{+1.00}_{-0.28}$\,kpc, or a Galactocentric 
distance of $12.58^{+0.85}_{-0.23}$\,kpc, and feature a metallicity ([Fe/H]$=-1.6^{+0.45}_{-0.30}$) and intermediate age ($9.5^{+3.0}_{-1.7}$\,Gyr). At the Galactic latitude of 31.788 deg, Kim\,3 is located $\sim8$\,kpc above the Galactic plane and therefore unlikely to be an old open (disk) cluster. 

The CMD of Kim\,3 in Figure~\ref{fig:CMD} appears to have a tight main-sequence, which implies the absence of binary stars with large mass ratios. This is in contrast with the observations of a strong anti-correlation between 
the fraction of binaries and the mass of the cluster~\citep{Milone2012,Milone2016}. Such a low binary fraction is more 
likely to be observed in the outer region of GCs as the binaries preferentially occupy the central region. Although the lack 
of a binary sequence in Figure~\ref{fig:CMD} might be the consequence of low number statistics, it implies that Kim\,3 possibly originated from the outskirts of a more massive GC undergoing tidal disruption in the gravitational field of the Milky Way. High precision photometry and proper-motion measurements will be able to test this hypothesis.

The half-mass relaxation time of Kim\,3 is estimated as $\sim50$\,Myr based on our measurements of 
the structural parameters in Section 3.3 and the equation (5) from~\cite{Spitzer1971}. As this time scale 
is significantly shorter than the observed age of Kim\,3 ($9.5^{+3.0}_{-1.7}$ Gyr), it is highly likely that 
the cluster has been dynamically relaxed for a long time and bears evidence of mass segregation. 
The left panel of Figure~\ref{fig:Contour} already gives an impression of mass segregation in Kim\,3 in the 
way that the majority of bright main-sequence stars between $20.0<r_{0}<21.5$ preferentially occupy the 
inner region of the cluster while the fainter, less massive MS stars mainly comprise the outer part of the 
cluster. The top panel of Figure~\ref{fig:MassSeg} shows the $r_{0}$ magnitudes of the 22 stars in the 
magnitude interval $19.5<r_0<23.5$ consistent 
with the main-sequence of the best-fit isochrone in the CMD within two half-light radii ($\sim1\farcm0$) as 
a function of radial distance from the center of the cluster. The lower panel shows the corresponding cumulative 
distributions for two different magnitude  intervals ($19.5<r_{0}<21.5$, $21.5<r_{0}<23.5$), corrected for incompleteness. Although it appears that the brighter (or more massive) main sequence stars are more common in the centre of Kim 3 than in its outskirts, a Kolmogorov-Smirnov (K-S) test implies that this seemingly mass-segregated state is not highly significant, yielding a formal probability of 87 percent that the two groups were sampled from populations with different parent distributions. This is possibly because of the relatively small sample sizes. The lack of a well defined center in the ultra-faint star cluster with an old stellar population also suggests that Kim\,3 
might have experienced substantial mass loss owing to tidal disruption in the gravitational field of the 
Milky Way \citep[see, e.g., discussion in][]{KimJerjen2015a}. The small overdensity 9.7\,pc away from Kim\,3 to the south, is likely a 
debris of the tidal disruption. We can use the centers of Kim\,3 and the overdensity as reference points 
to determine the associated great circle. Taking it as an approximation for the orbital path of Kim\,3,
we find that the two globular clusters $\omega$\,Centauri ($d_{gc}=6.4$\,kpc) and 
NGC5286 ($d_{gc}=8.4$\,kpc), which are $\sim16.9^\circ$ and $\sim21.2^\circ$ away from Kim\,3,
are only $2.5^\circ$ and $0.2^\circ$ away from that great circle. In this context it is further interesting to note that these systems are among the few MW globular clusters showing internal variations in metals~\citep{Marino2015}, which led to the hypothesis they are surviving remnants of tidally disrupted dwarf galaxies. Kim 3 may have originated from a more massive stellar system that also 
hosted NGC5286 or Omega Centauri. Future radial velocity and proper motion measurements will help to test this idea.

\acknowledgements{The authors thank Tammy Roderick, Gyuchul Myeong and Alfredo Zenteno for 
their assistance during the DECam observing run, and Aaron Dotter for his help with the 
Dartmouth isochrones. We also thank the referee for the helpful comments and suggestions on the original manuscript. HJ and GDC acknowledge the support of the Australian Research Council through 
Discovery project DP150100862.

This paper makes use of data from the AAVSO Photometric All Sky Survey, whose funding has been provided by the Robert Martin Ayers Sciences Fund. This research made use of Astropy, a community-developed core Python package for Astronomy~\citep{astropy}, and Matplotlib library~\citep{matplotlib}.

This project used data obtained with the Dark Energy Camera (DECam), which was constructed by the Dark Energy Survey (DES) collaborating institutions: Argonne National Lab, University of California Santa Cruz, University of Cambridge, Centro de Investigaciones Energeticas, Medioambientales y Tecnologicas-Madrid, University of Chicago, University College London, DES-Brazil consortium, University of Edinburgh, ETH-Zurich, Fermi National Accelerator Laboratory, University of Illinois at Urbana-Champaign, Institut de Ciencies de l'Espai, Institut de Fisica d'Altes Energies, Lawrence Berkeley National Lab, Ludwig-Maximilians Universitat, University of Michigan, National Optical Astronomy Observatory, University of Nottingham, Ohio State University, University of Pennsylvania, University of Portsmouth, SLAC National Lab, Stanford University, University of Sussex, and Texas A$\&$M University. Funding for DES, including DECam, has been provided by the U.S. Department of Energy, National Science Foundation, Ministry of Education and Science (Spain), Science and Technology Facilities Council (UK), Higher Education Funding Council (England), National Center for Supercomputing Applications, Kavli Institute for Cosmological Physics, Financiadora de Estudos e Projetos, Funda\c{c}\~ao Carlos Chagas Filho de Amparo a Pesquisa, Conselho Nacional de Desenvolvimento Cient\'i­fico e Tecnol\'ogico and the Ministério da Ci\^encia e Tecnologia (Brazil), the German Research Foundation-sponsored cluster of excellence "Origin and Structure of the Universe" and the DES collaborating institutions.}


\end{document}